\documentclass{interact}

\usepackage{natbib}
\bibpunct[, ]{(}{)}{;}{a}{}{,}

\usepackage{amsmath,amssymb}
\usepackage{graphicx}
\usepackage{etoolbox}      
\usepackage{hyperref}

\graphicspath{{./}{fig/}}

\let\bbordermatrix\bordermatrix
\patchcmd{\bbordermatrix}{8.75}{4.75}{}{}
\patchcmd{\bbordermatrix}{\left(}{\left[}{}{}
\patchcmd{\bbordermatrix}{\right)}{\right]}{}{}

\newcounter{algocntr}
\labelformat{algocntr}{Listing~#1}
\newenvironment{algo}[1]{%
 \refstepcounter{algocntr}\label{#1} \noindent \small \begin{tabular}{llllll}
}{%
 \hline \\ \end{tabular}
}
\newcommand{\alghdr}[1]{\hline \multicolumn{6}{l}{%
 \vbox{ \vspace*{6pt} \hbox{Listing \thealgocntr: \text{#1}} \vspace*{0pt}} %
 }\\ \hline}
\newcommand{\algrow}[2][]{#1 & \multicolumn{5}{l}{#2} \\}
\newcommand{\algrowi}[2][]{#1 && \multicolumn{4}{l}{#2} \\}
\newcommand{\algrowii}[2][]{#1 &&& \multicolumn{3}{l}{#2} \\}
\newcommand{\algrowiii}[2][]{#1 &&&& \multicolumn{2}{l}{#2} \\}
\newcommand{\algrowiv}[2][]{#1 &&&&& #2 \\}

\newcommand{\sets}[1]{\mathcal{#1}}
\newcommand{\rpkg}[1]{{\em #1}}
\newcommand{\sizeof}[1]{{\lvert #1 \rvert}}
\newcommand{\func}[1]{{\sf #1()}}

\begin{document}

\title{Using Spacing to Detect Multi-Modality}
\author{
  \name{Greg~Kreider}
  \thanks{CONTACT Greg Kreider.  Email: gkreider@primachvis.com}
  \affil{Primordial Machine Vision Systems, Lyndeborough, NH, USA} }

\maketitle

\begin{abstract}

The spacing of a unimodal variate resembles a `U' with a flat bottom and
rapidly increasing values in the tails.  When combined into multi-modal
setups, the transition between variates into their tails forms local
increases in the spacing.  These peak features signal the data is
multi-modal while the flats locate the modes.  We develop tests of the
features, including parametric models based on an assumed null univariate
distribution, the number of runs or longest run in the signed difference
of the interval spacing, and feature reconstruction by means of
permutations of the runs or bootstrap samples from a pool based on the
difference of the signal.  We also try combining existing changepoint
detectors to look where the behavior of the spacing changes.  This paper
describes the processing of the spacing and models and tests in full,
gives examples of their use, and summarizes the performance in general.

\end{abstract}

\begin{keywords}
spacing; multi-modality ; parametric models ; runs tests ; changepoint detectors
\end{keywords}
\begin{amscode}
AMS 62E10 ; AMS 62G30
\end{amscode}


Imagine a histogram built from draws from two normal variates, with two
rounded hills in the combined density centered at the means and a sag
in-between.  The separation within a bin is inversely proportional to
the count, assuming points are scattered evenly over its width.
Spacing, the difference between adjacent order statistics, is another
way of expressing this separation.  It reflects the modality of the
data, with consistent, similar values near the mode at the center of
the hill and a local rise at the transition in-between.

Analyzing such features is difficult.  The spacing has a high variance
and its unevenness obscures the changes.  Smoothing by some means seems
necessary, and we begin our study by applying a low-pass filter and
developing models of the feature's characteristic value --- the height of
peaks or length of flats --- assuming a null distribution.  But this is
equivalent to kernel density estimation and bump hunting in the result,
and so we look further. Taking the spacing over larger intervals than
adjacent order statistics is a poor man's filter, equivalent to a
rectangular or running mean kernel, which does not suppress sidelobes
much.  It passes more of the high frequency components, meaning the
interval spacing will be rough.  We can use this to look at runs in
its signed difference, or series of increasing, decreasing, or tied
values.  Combinatorial analysis tells us the expected number of runs
within a feature, or we can model the data with a Markov chain to
find the probability of the longest run, or we can reconstruct the
feature from permutations of its runs to determine its likelihood.
Extending the last process, we adapt bootstrap sampling to reconstruct
features from a pool of the differences in either spacing as a fifth
test.  Finally, we look at changepoint detectors to find behavioral
changes in the raw spacing, without any filtering.

This work is mostly the application or extension of known techniques
to the problem, because a theoretical approach is unlikely to provide
concrete results, given the generality of the data and intractability
of the math.  The models are new, as is the Markov chain model of the
run length.  The results show that spacing can be used to identify
multi-modal data and to locate the transitions between modes.  This
paper describes the existing theory of spacing and defines the
feature detectors and the tests for their significance.  It provides
a summary of the performance and discussion of the benefits and
limitations of the approaches, but separate papers present a more
thorough evaluation of the tests \cite{kreider25b} and the reference
implementation in the \rpkg{Dimodal} package for R \cite{kreider25c}.
The evaluation includes variations of a bi-modal setup to determine
the sensitivity and accuracy of the features and their stability and
repeatability over many draws of a large set of samples from the
literature. \cite{kreider24} is a detailed write-up of the
development of this work, including the data and decisions behind
the models and a full description of the evaluation results,
although some of these no longer apply to the final versions of the
detectors and tests in \rpkg{Dimodal}.

\section{Spacing and Modality}
\label{sec:Di}

In general the spacing of a variate has the form of a `U' with a
very flat bottom \cite[Figure~1]{kreider23a}.  The arms represent
the initial and trailing tails: the variate's distribution function
is sigmoidal or logistic in shape, subject to the variate's
symmetry, and the separation between uniformly placed quantiles
grows quickly in the asymptotes.  The center comes from the
transition region of the sigmoid, which is roughly linear,
producing similar separations.  The mode, located at this
transition, will therefore see consistent spacing, while away
from it the spacing increases.  The transition depends on the
variate's scale parameters, and the spacing within the mode
will reflect them.

This behavior extends to multi-modal situations.  So long as the
transitional zone in each distribution function is distinct, the mode
will generate data with consistent spacing that depends on the scale
parameters of the variates.  The length of this region depends on the
distribution and size of the draw.  Between them the spacing will
increase, although not as strongly as in the tails.  The size of the
increase will depend on the location parameters.  We therefore have
two features in the spacing that indicate multi-modality: flat or
consistent regions that cover the mode, and local peaks at the
anti-modes.

For example, the expected spacing of a logistic variate is
\cite[(17)]{kreider23a}
\begin{equation} \label{eq:edi.logis}
E\Bigl\{ D_{i,logis} \Bigr\} = \frac{\sigma n}{(i-1) (n-i+1)}
\end{equation}
where $ \sigma $ is the scale, $ n $ the draw size, and $ i $ an index
between 2 and $ n $; the spacing is taken at the upper point of the pair.
A third of the points stay within 10\% of the mode at $ i = n/2 $.  The
spacing at the first and last points is $ n / 4 $ times larger, a factor
of 25 for 100 points.  For an exponential variate the flat extends over
a tenth of the sample and the tail is $ 99 \times $ larger.

The theory of spacing is well-developed.  Using the notation of Pyke
\cite{pyke65} with the indexing above, the spacing is
$ D_{i} = T_{i} - T_{i-1} $ with $ T $ the order statistics.  Its
density is
\begin{equation} \label{eq:fdi}
f_{D_{i}}(y) = \frac{n!}{(i-2)! (n-i)!}
  \int_{-\infty}^{\infty}
    \left\{ F(x) \right\}^{i-2} \left\{ 1 - F(x+y) \right\}^{n-i}
    f(x) f(x+y) ~ dx
\end{equation}
for a variate with density function $ f $ and distribution or c.d.f. $ F $.
The expected spacing is the first moment
\begin{equation} \label{eq:edi}
E\Bigl\{ D_{i} \Bigr\} = \int_{0}^{\infty} y f_{D_{i}}(y) ~ dy
\end{equation}
and can be solved in a closed form for only a few variates
\cite{pyke65, kreider23a}.  These equations also apply to multi-modal
setups, where the density and distribution functions are the sums of the
individual variates weighted by the draw size.  Except for uniform
samples, we cannot find expressions for the expected spacing because the
factors with powers of the combined distribution do not separate.
Attempts to model the setup with a Taylor expansion yielded neither
accurate approximations nor insight into how the spacing responds,
especially between modes. We cannot study multi-modal spacing, except
numerically.

\begin{figure}
\centering
\begin{minipage}[t]{\textwidth}
\includegraphics[width=\textwidth]{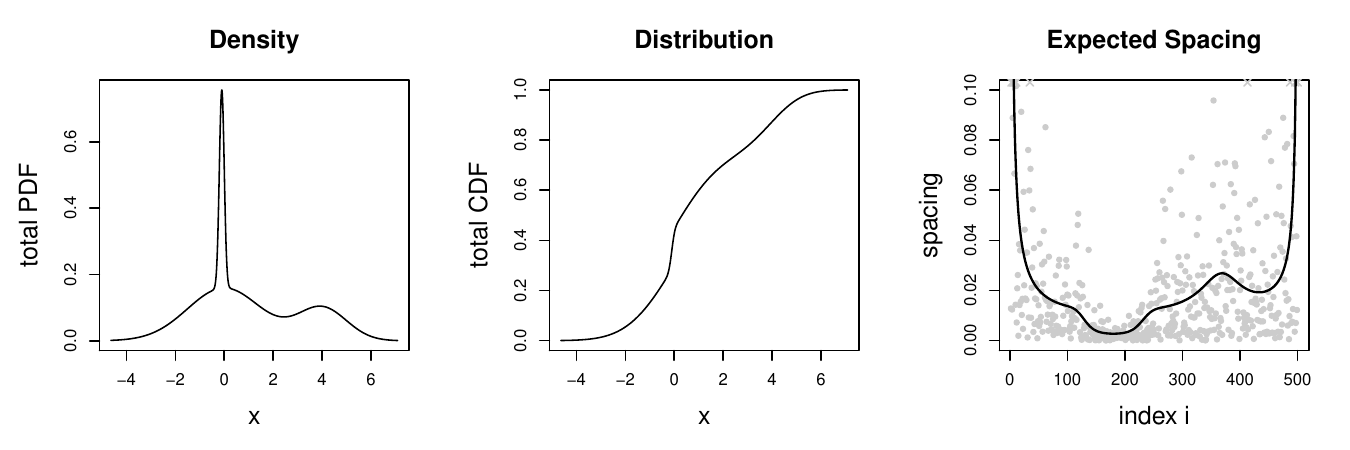}
\caption{\label{fig:edi} Expected spacing in tri-modal setup
 \eqref{eq:trimode}.}
\end{minipage}
\end{figure}

Figure~\ref{fig:edi} shows an example with three normal variates,
\begin{equation} \label{eq:trimode}
75 \times N(-0.1, 0.1), \quad 300 \times N(0, 1.5),
  \quad 125 \times N(4.0, 1.0)
\end{equation}
where $ N(\mu,\sigma) $ is a Gaussian with mean $ \mu $ and standard
deviation $ \sigma $.  The tight width of the first variates produces
a small spacing between indices 150 and 200 in the right graph.  In
the total distribution function the steep increase around $ x = 0 $
from the first variate breaks the linear section between $ -2 $ to
$ 2 $ from the second, which forms shoulders or shelves to either
side.  The transition from flat to shoulder is smooth in this setup,
but shifting the first mean to $ -0.6 $ would create a dip in the
density to its right, which would raise a peak in the spacing.  The
density has a minimum at $ +2.442 $ between the second and third
modes.  The peak in the expected spacing at index $ i = 368 $
corresponds to $ +2.477 $.  The overall spacing still follows the
`U' and the steep arms blend with the third variate and round off
its flat.  The growth in the arms is strong compared to the global
minimum of 0.003, to 0.48 at the first point and 0.35 at the last.
Four points in the leading tail and three in the trailing fall
outside the graph.

Spacing from a single draw will be very rough, with a variance
comparable to the square of the expected value.  The dots in the
third graph show a sample draw, with points outside the $ y $
limit marked at the graph's top with crosses.  The local increase
at index 375, with outliers at 353 and 412, and the tight first
draw are clear, but the background level beyond index 250 is high
and a spurious increase lies at index 110.  Some smoothing is
needed, and a finite impulse response (FIR) filter is ideal.  Our
experience during model development agrees with the conclusion of
\cite{harris78} and its equivalent noise bandwidth metric: a
Kaiser filter designed following
\cite[Section 7.4.3]{oppenheim89} performs best.  We choose not
to handle the case where the filter extends beyond the data,
ignoring points within half the filter size of the start and end.
While this helps to screen the tails or the arms of the `U', it
can also lose features at the edges.  An advantage of low-pass
filtering is its ability to smooth over the steps between
discrete or heavily quantized data, for example when it is
recorded with a limited number of decimal points, as we will see
in the examples. Without filtering such spacing consists of
single point impulses at the steps where the values change and
is zero between.  With filtering the signal responds to the
density of steps, and peak and flat features can appear in it.

An alternative that has been used in modality analysis is using
spacing taken over larger gaps than adjacent order statistics.
Call this the ``interval spacing'', where
$ D_{i,w} = T_{i} - T_{i-w} $ for $ n \ge i > w $ following the
notation above.  \cite{grenander64} and \cite{venter67} use it to
estimate the position of a mode, where the interval width $ w $ is
chosen depending on the curvature of the distribution about the
mode.  \cite{duembgen08} defines a statistic combining interval
spacings to test for regions of non-zero slope in the data.  Its
density
\begin{align} \label{eq:fdiw.gen}
f_{D_{i,w}}(y) & =
\begin{aligned}[t]
 S_{1} \int_{-\infty}^{\infty} &
 \left\{ F(x) \right\}^{i-w-1} \left\{ F(x+y) - F(x) \right\}^{w-1}
 \left\{ 1 - F(x+y) \right\}^{n-i} \\
 & \qquad \cdot f(x) f(x+y) ~ dx
\end{aligned} \nonumber \\
S_{1} & = \frac{n!}{(i-w-1)! (w-1)! (n-i)!} 
\end{align}
follows from the joint density of two order statistics,
\cite[(8) and (31), with $ r = i-w $ and $ r' = i $]{wilks48} or
\cite[(2.4)]{pyke65}.  There are solutions of this and its first
moment for uniform, exponential, and logistic variates, but they
involve series and are not particularly helpful, needing
high-precision math libraries to counteract the scaling of
$ S_{1} $ for $ i $ near the middle of large $ n $.  More
importantly, the interval spacing is the sum of individual
spacings over the width,
\begin{align}
D_{i,w} & = T_{i} - T_{i-w} \nonumber \\
 & = (T_{i} - T_{i-1}) + (T_{i-1} - T_{i-2}) + \ldots 
     + (T_{i-w+2} - T_{i-w+1}) + (T_{i-w+1} - T_{i-w}) \nonumber \\
 & = \sum_{j=0}^{w-1} D_{i-j} 
\end{align}
But this is just a rectangular FIR filter run over the spacing, without
normalizing the sum by the width so that it appears to amplify
differences.  The interval spacing is a low-pass version of the spacing.
The rectangular filter, also called a running mean, has a wide mainlobe
but also poor sibelobe suppression.  The first allows for smaller
intervals.  The second means that high frequency components will pass
through as roughness or ringing, and although this makes peak detection
more difficult, it may help preserve some small features, especially
close to the tails.

\section{Feature Detectors}
\label{sec:feat}

Features are defined in terms of the range of a signal, not
necessarily the spacing.  Identifying the peaks and flats is
simple, but an important part of the two detectors is defining
which to keep for analysis.  Small peaks, perhaps on the shoulders
of larger, and short flat sections can be ignored.

\begin{figure}
\centering
\begin{minipage}[t]{0.66\textwidth}
\includegraphics[width=\textwidth]{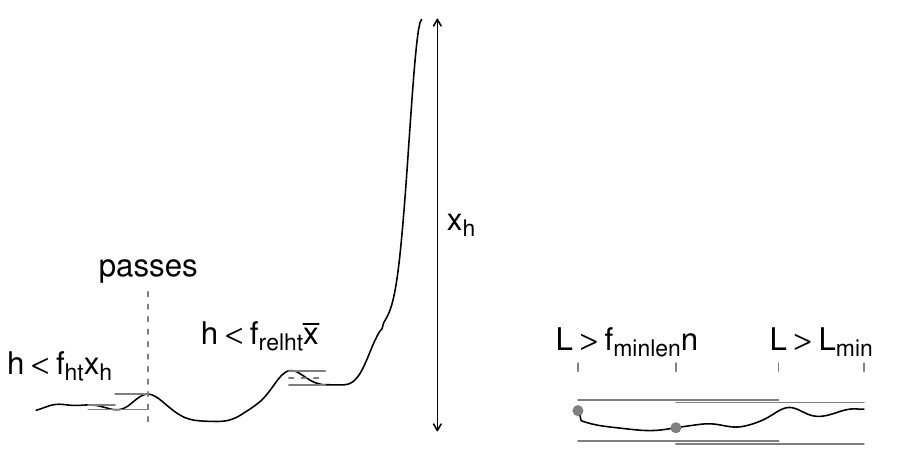}
\caption{\label{fig:feat} Merged peak (left) and overlapping flat (right)
 requirements.}
\end{minipage}
\end{figure}

Local extrema are points that are larger or smaller than both
neighbors. To accommodate noise and numerical precision,
nearly-equal points collapse to one before marking them.  This
turns square waves into alternating high and low points, for
example.  To ignore minor features or subsidiary peaks, a local
maximum merges into its neighbor if its height above the shared
minimum is small.  Said differently, shallow minima are deleted.
The height must be greater than a fraction $ f_{ht} $ of the
total data range or $ f_{relht} $ of the average value of the
peak and the minimum.  The first imposes a global, data-dependent
threshold, the second is local and relative.  The left trace of
Figure~\ref{fig:feat} shows two peaks that do not meet these
requirements, while the peak between them does.  The rejected
peak to the left would merge into the one that passes, the right
into the steep drop.  The height is the difference between solid
bars, and the average value is marked with a dashed bar.  Default
values of the criteria are $ f_{ht} = 0.05 $ and
$ f_{relht} = 0.15 $.  The compaction of nearly-equal points
uses the same relative difference with a much tighter threshold,
$ f_{h,tie} = 0.001 $.  The detector places no restrictions on
the placement of peaks, only their heights.

Flats are traditionally defined as successive points that fall within
a ripple specification, a fraction $ f_{ripple} $ of the data range
and centered about a source point.  The characteristic value is its
length $ L $ since the ripple fixes the height.  The detector checks
all data points, keeping the longest flat at each if it meets a
minimum length requirement, either an absolute value $ L_{min} $ or a
fraction of the data $ f_{minlen} $.  Two flats may overlap if the
segment not covered by the longer meets the requirement.  The right
side in Figure~\ref{fig:feat} shows two overlapping flats bounded
vertically by the ripple range.  Either extension betyond the common
segment meets the length requirement.  Default values of the
parameters are $ f_{ripple} = 0.05 $, $ L_{min} = 30 $, and
$ f_{minlen} = 0.05 $.

Appendix~\ref{app:detect} presents pseudo-code for both detectors.

\section{Parametric Models}
\label{sec:models}

The models capture the frequency of a feature's characteristic value ---
peak height or flat length --- based on univariate draws from a base
distribution.  Their development started with a quick evaluation to
choose the variate and default filter, followed by a very large number
of simulations to get robust quantiles out to 0.9995 over the parameter
space, and finished with regressions to fit the values.  The models
use two parameters, the size $ n $ of the draw, varying from 50 to
500 points, and the size $ f_{lp} $ of the low-pass filter kernel as a
fraction of the data, varying between 0.05 and 0.40, and collect the
critical values at quantiles $ q $ ranging from 0.90 to 0.9995, which
covers the range of interest for significance testing.  This process
is an exercise in developing good fits between the parameters and
characteristic values; there is no theoretical basis for the choices
made. We could, for example, attempt to use one of the Brownian
excursion variants to describe a peak, but they rely on the
underlying process being normal, which the spacing is not.

The setup for the peak height model is clear.  Of thirteen variates
tried for the null distribution, two can be ruled out.  Uniform draws
produce many large peaks and its 0.90 quantile height lies beyond the
0.9995 quantile of all other distributions.  Beta variates match the
others slightly better, but its 0.90 quantile still corresponds to
their 0.999 level.  The remaining distributions do cluster together,
and the asymmetric Weibull, with scale $ a = 2 $ and shape $ b = 4 $,
is the conservative choice.  Its $ q = 0.95 $ heights are no worse
than those at $ q = 0.99 $ for $ t $ and logistic draws.  As
expected, the Kaiser filter performs best, producing the fewest and
smallest peaks.  The other filters generate critical values that are
roughly a constant multiple of the Kaiser's for the quantiles of
interest between 0.90 and 0.999, and we can scale the base model to
accommodate a different filter choice.  Gaussian and Blackman
kernels behave very similarly, produce 30\% larger peaks than the
Kaiser, and have the most variation over the quantile range.  

To account for scaling effects within the null distribution we take
as the peak height the largest of the rises to the left and right
neighboring minima and divide by the standard deviation
$ \sigma_{lp} $ of the draw after filtering;
\begin{equation} \label{eq:hpeak}
h_{peak} = \max(h_{l}, h_{r}) / \sigma_{lp}
\end{equation}
A preliminary fit shows this varies with $ \log n $ but that there is no
good combination with $ f_{lp} $ or $ q $.  We instead model the heights
with a distribution function and find the Wald or inverse Gaussian fits
best, of ten variates evaluated.  This happens to describe Brownian
excursions, although again we say that there is no theoretical reason
for making this link.  The Wald density and distribution functions take
two parameters, a location $ \mu $ and shape $ \lambda $,
\begin{align} \label{eq:Fwald}
 f_{Wald}(x, \mu, \lambda) &
   = \sqrt{\frac{\lambda}{2 \pi x^{3}}} ~
   e^{-\lambda (x - \mu)^{2} / 2 \mu^{2} x} \nonumber \\
 F_{Wald}(x, \mu, \lambda) &
  = \Phi\left( \frac{x}{\mu - 1} \sqrt{\lambda / \mu} \right)
  ~-~ e^{2 \lambda / \mu} ~ \Phi\left( \frac{x}{\mu + 1} \sqrt{\lambda / \mu} \right)
\end{align}
The quantile of the height is
\begin{align} \label{eq:ppeak}
q & = F_{Wald}\left( 10^{((h_{peak} / f_{filter}) - b_{adj}) / m_{adj}},
                     \mu, \lambda \right) \\
\intertext{with}
b_{adj} & = -0.2305 + 11.8716 f_{lp} - 46.9360 f_{lp}^{2}
               + 85.0096 f_{lp}^{3} \nonumber \\
m_{adj} & = 4.2412 - 7.2054 f_{lp} + \left( 0.1547 / f_{lp} \right) \nonumber \\
\mu & = (5.8158 + 2.4152 \log f_{lp})
        - (1.9704 - 1.0131 \log f_{lp}) \log n \nonumber \\
\lambda & = (-2.0204 + 49.7357 f_{lp}) + (2.6034 - 19.5195 f_{lp}) \log n
 \nonumber
\end{align}
The scaling of the height is polynomial in $ f_{lp} $ and the draw
size enters in the Wald parameters, also in combination with the
filter size.  Inverting, the critical value is
\begin{align} \label{eq:cvpeak}
h_{peak} & = f_{filter} \left( b_{adj}
               + m_{adj} \log F^{-1}_{Wald}(1-q, \mu, \lambda) \right)
\end{align}
$ f_{filter} $ in \eqref{eq:ppeak} refers to the scale factor from the
filter kernel choice and is 1.0 for the Kaiser.  The arbitrary nature
of the parameters and linear adjustment of the height is clear.  The
actual filter width $ f_{lp} n $ plays no role in this model.

The choice of base distribution for flats is not clear, but the model
itself is simpler.  Flats are the inverse of peaks, and the variates
that generate peaks do not create many flats.  We see a wider spread
in the lengths. Uniform and beta variates now generate too few flats,
and F, exponential, and Wald too many.  The Gumbel and asymmetric
Weibull null distributions bracket the remaining critical values,
with the Weibull the more liberal in the sense of accepting shorter
features. Its critical lengths are 60\% of the Gumbel.  A logistic
null distribution is a compromise, in-between these two, and
therefore is chosen for the default model.  For example, at
$ q = 0.95 $ the length of a Weibull flat is 35, of a logistic 52,
and of a Gumbel 68.  At $ q = 0.99 $ these increase to 49, 64, and
94.  The variation between distributions and quantiles is strong,
with much less overlap than seen with the peak heights.  

The choice of filter for the flat model is less important.  The Kaiser
is again conservative, passing at the 0.95 level what other filters do
at 0.99.  Unlike peaks, other filters do not simply scale off the
Kaiser critical values and separate simulations and fits are needed
for each. The combination of three base distributions and six filter
options substantially increases the number of simulations needed.

On the other hand, the models are straightforward and have smaller
residual errors than the height models.  The regression involves 24 terms,
\begin{equation} \label{eq:pflat.len}
l_{flat} \propto (c_{flp} + f_{lp}) \times (c_{n} + n + n^{2}) \times
                 (c_{q} + q + q^{2} + \log q/(1-q))
\end{equation}
where the $ c $ are constants.  Half  of the coefficients are small and
could be dropped while still keeping a reasonable fit, but the subset
changes with the combination of base distribution and filter.  We always
use the full set. We could also use only the actual kernel size
$ f_{lp}\,n $ by deleting the separate $ f_{lp} $ term, with less than
a 1\% increase in the total residuals, but for consistency with the
peak model we do not.  Table~\ref{tbl:flatmodel} has the model
coefficients for the logistic base distribution with Kaiser filter.
A numeric root finder inverting \eqref{eq:pflat.len} determines the
quantile of a length.

\begin{table}
\begin{center}
\caption{\label{tbl:flatmodel} Flat Model For a Logistic Base Distribution
  and Kaiser Filter}
\begin{tabular}{ccccccl}
$ c_{n} c_{lp} $&$ n $&$ n^{2} $&$ f_{lp} $&$ f_{lp} n $&$ f_{lp} n^{2} $& \\
{\small   895.07 }&{\small -6.4915 }&{\small  0.050202 }&{\small -2914.0 }&
  {\small  39.755 }&{\small -0.12859 }& $ c_{q} $ \\
{\small -1945.1  }&{\small 14.621  }&{\small -0.10957  }&{\small  6365.8 }&
  {\small -87.586 }&{\small  0.28017 }& $ q $ \\
{\small  1068.6  }&{\small -8.4033 }&{\small  0.060307 }&{\small -3524.0 }&
  {\small  49.338 }&{\small -0.15443 }& $ q^{2} $ \\
{\small -0.80887 }&{\small 0.078324 }&{\small -6.8829e-5 }&{\small 3.8649 }&
  {\small -0.16948 }&{\small 0.19127e-4 }& $ \operatorname{logit}(q) $
\end{tabular}
\end{center}
\end{table}


The peak height model is slightly biased towards accepting features and
should be tested at the 0.01 level to control false positives at the
0.95 quantile.  The logistic flat length model is conservative and can
be used at the 0.05 level.

\section{Runs Tests}
\label{sec:runs}

To avoid using a null distribution, however conservative it may be,
we looked at non-parametric tests to evaluate the peaks.  Analyzing
sequences of symbols provides several such approaches, be it
combinatorial counting of cards, rank comparison tests such as the
Wilcoxon rank-sum statistic, or trend analysis in economic data
reduced to ``increasing'' and ``decreasing'' symbols.  This latter
is particularly interesting as we can apply it to the spacing,
looking at runs in the sign $ -1 $, 0, or $ +1 $ of the difference
between adjacent points, where tied values are possible with
quantized data.  The interval spacing is the appropriate source of
data, as the low-pass filter smooths away local differences and
the raw spacing has no significant runs.

The first runs test was done on continuous data \cite{wallis41a} so
there were only two symbols, $ -1 $ and $ +1 $; the test used a modified
chi-squared distribution determined empirically because the amount of
data adds a condition on the total length of all runs that cannot be
captured analytically.  The test is rather crude because the expected
number of runs drops off very quickly and in practice only those of
length 1, 2, or 3 are counted.  Wald and Wolfowitz built a summary
statistic using all the runs in a combinatorial counting of two
symbols \cite{wald40}.  Kaplansky and Riordan extended this to any
number of symbols, showing that runs count $ U $ is distributed
normally with mean and variance given in
\cite[(12) and (13)]{kaplansky45}.  For our problem their equations
become
\begin{align} \label{eq:krtest}
E\{U\} & = 1 + \frac{2 \alpha_{2}}{\alpha_{1}} \nonumber \\
V\{U\} & = \frac{2 \alpha_{2} (2 \alpha_{2} - \alpha_{1})
                 - 6 \alpha_{1} \alpha_{3}}{\alpha_{1}^{2} (\alpha_{1}-1)}
\end{align}
where
\begin{alignat*}{2}
\alpha_{1} & = \sum_{i=i}^{s} a_{i} 
           && = a_{1} + a_{2} + a_{3} \\
\alpha_{2} & = \sum_{i=1}^{s-1} \sum_{j=i+1}^{s} a_{i} a_{j}
           && = a_{1} a_{2} + a_{1} a_{3} + a_{2} a_{3} \\
\alpha_{3} & = \sum_{i=1}^{s-2} \sum_{j=i+1}^{s-1} \sum_{k=j+1}^{s}
   a_{i} a_{j} a_{k}
           &\quad& = a_{1} a_{2} a_{3}
\end{alignat*}
in general and specifically for three symbols.  Without ties, where one
$ a_{i} $ is 0, this simplifies to the Wald and Wolfowitz result.  The
test considers each peak separately between its bounding minima and
counts the symbol populations $ a_{i} $ and the number of runs within.
\cite{shaughnessy81} generalizes the problem further, to consider the
probability of runs of a specific length, finding recursive expressions
for the runs count, but \eqref{eq:krtest} suffices for our application.

We want to account for the correlations between data points arising
from the common portion of successive intervals, however.  We
introduce a new test on the length of the longest run based on a
Markov chain model that captures such interdependencies.  Let
$ {\bf T } $ be the symbol transition matrix, estimated normally from
the frequency of symbol pairs in the complete data.  Divide this
matrix into diagonal elements $ {\bf A} $ that advance the run length
for each symbol and off-diagonal elements $ {\bf B} $ that switch
symbols and begin a new run; $ {\bf T} = {\bf A} + {\bf B} $.  Each
matrix has size $ s \times s $, where $ s $ is the number of symbols.
Create a new transition matrix $ {\bf R} $ capturing runs of length
$ L $
\begin{equation}
{\bf R} =
\bbordermatrix{ %
 \overset{\text{next}}{\text{\scriptsize length}} & \text{\scriptsize 1} &
   \text{\scriptsize 2} & \text{\scriptsize 3} & \text{\scriptsize 4} & &
   \text{\scriptsize L} & \text{\scriptsize L+1} \cr
 & {\bf B} & {\bf A} & 0 & 0 & & 0 & 0 \cr
 & {\bf B} & 0 & {\bf A} & 0 & & 0 & 0 \cr
 & {\bf B} & 0 & 0 & {\bf A} & & 0 & 0 \cr
 & & & & & \ddots & & \cr
 & {\bf B} & 0 & 0 & 0 & & {\bf A} & 0 \cr
 & {\bf B} & 0 & 0 & 0 & & 0 & {\bf I} }
\end{equation}
The identity matrix $ {\bf I} $ absorbs any runs longer than $ L $.  The
probability that this happens is
\begin{equation} \label{eq:plongrun}
P\left\{ \text{len} > L \right\} = {\bf w} ~ {\bf R}^{N}[1,L] =
  {\bf w} ~ {\bf r}_{1,L,N}
\end{equation}
$ {\bf r} $ is the upper right $ s \times s $ sub-matrix in row 1 and
column $ L $ after running the chain over $ N $ steps.  $ {\bf w} $ is
an $ s \times 1 $ initial weighting vector, which can be the stationary
state of $ {\bf T} $ in general or a unit vector if starting from a
known single state.  We assume the first run starts with the feature,
or would otherwise have to include other rows and a weighting factor
for the in-progress length at the start.  The probability of the
longest run matching $ L $ is
\begin{equation} \label{eq:maxrun}
P\left\{ \text{len} = L \right\}
  = {\bf w} \, \left( {\bf r}_{1,L-1,N} - {\bf r}_{1,L,N} \right)
\end{equation}
``Running the chain'' means calculating $ {\bf R}^N $, which gives a
recursion for the upper right element
\begin{equation} \label{eq:recuralt}
{\bf r}_{1,L+1,N} =
  {\bf A}^{L} + \sum_{j=1}^{L} {\bf A}^{j-1} ~ {\bf B} ~ {\bf r}_{1, L+1, N-j}
\end{equation}
with seed
\begin{equation*} \label{eq:recuraltst}
{\bf r}_{1,L+1,n} = \left\{
\begin{aligned}
0 & & \qquad n < L \\
{\bf A} & & n = L
\end{aligned}
\right.
\end{equation*}
Since $ {\bf A} $ is diagonal the powers in \eqref{eq:recuralt} are
done by raising its elements to that power, without multiplying
matrices.  The recursion therefore is $ O(LN) $ in matrix
operations.  It could also be calculated over the rows of the last
column of $ {\bf R} $, which is more efficient since $ L < N $.

A third non-parametric approach is to look at how runs combine to form
a peak, using its height as a test parameter.  Here we use both the
run length and direction with the reconstructed feature $ x_{perm} $ a
cumulative sum of the product of the two.  If the peak is made of
two long runs, one increasing and one decreasing, plus a large number
of small runs in either direction, then a height matching the length
of the long runs will not be unusual.  Indeed, when there are only
the two long runs it will be inevitable.  But peaks formed from more
and smaller runs must separate, with short drops interspersed among
moderate rises to the left of the peak and short rises interrupting
moderate falls during the decline.  Let the reconstructed height
$ h_{perm} $ be the largest difference in the sum to the first or
last point,
\begin{equation} \label{eq:pkht}
h_{perm} = max(x_{perm}) - min(x_{perm}[1], x_{perm}[n])
\end{equation}
This definition mirrors the actual height of peaks above the minima.
It removes permutations that create minima or valleys.  Then the
feature's quantile is the fraction of all permutations of its runs
that produce peaks that meet its height.  That is,
\begin{equation} \label{eq:pperm}
q\{\text{peak height} < H\} =
 \left( \#\{ h_{perm} < H \} + \#\{ h_{perm} = H \} / 2 \right) / N_{perm}
\end{equation}
with $ \#\{\} $ the counting function.  Because the runs have integer
length the heights are discrete, which creates coarse steps in the
counting function.  The second term in \eqref{eq:pperm} is a discrete
mid-quantile approximation that corrects for the step \cite{ma11}.
It estimates the quantile better, but does not remove the step size;
the test's resolution can be poor, especially in the tails of the
draw which is the most interesting for significance testing. A
restriction on the permutations is that adjacent runs may not have
the same sign, otherwise they would form a longer run that does not
exist in the feature.  The test, as described in
Appendix~\ref{app:excur}, does the permutation in two phases, first
creating an alternating sequence of rising and falling symbols and
then assigning lengths within each sign by a shuffle.  The
restriction is substantial, with fewer than 5\% of all possible
permutations surviving; the fraction shrinks as the number of runs
increases.  Still, with more than eight runs there are too many
permutations to check exhaustively and the test must be done by
sampling.  $ N_{perm} = 5000 $ draws is enough for a stable
distribution for \eqref{eq:pperm}.

The runs tests are performed on each peak found in the interval
spacing, bounded by the minima to either side.  The signed difference
over the range is passed to the runs detector, and the probability of
the feature calculated either from the number of runs compared to the
expected value based on the symbol distribution, or the longest run
based on the overall transition rates between symbols, or the
permutations.

The count and length tests perform the same.  Judged at the 0.05
level they have a 20\% misclassification rate when the data is
not multi-modal.  At the 0.01 level this false positive rate
disappears.  The height permutation test is not decisive and
responds to marginal peaks, so a tight passing level of 0.005 is
needed to reduce false positives. Even so the test is biased
towards accepting peaks.

\section{Bootstrap/Excursion Tests}
\label{sec:excur}

We can extend the permutation test to evaluate signals in general by
using the point-by-point difference --- in other words, its first
derivative --- as the pool to draw from, rather than runs.  When
applied to the spacing this amounts to a second derivative of the
order statistics.  Sampling is done with replacement, making this
a bootstrap test, and the signal $ x_{excur} $ reconstructed by a
cumulative sum of the drawn differences.  We call this an excursion
test, comparing the feature height against repeated trials of the
bootstrap.  Like the permutation test the quantile of a peak is the
fraction of the trials with smaller or matching heights. Again the
simulated height is defined as the rise above the start and end of
the excursion, and again we make a mid-quantile correction if the
data is discrete. But the test can also be used to evaluate flats
by counting the trials whose reconstructed heights, here the total
range to match the ripple parameter, are larger.  That is,
\begin{equation}
h_{excur,flat} = \max(x_{excur}) - \min(x_{excur})
\end{equation}
and
\begin{equation}
q\{\text{flat height } < H\} =
 \left( \#\{ h_{excur} > H \} + \#\{ h_{excur} = H \} / 2 \right) / N_{excur}
\end{equation}
This essentially swaps the roles of the feature parameters, treating
the length as given and the height as the dependent variable.

The initial and final tails of the `U' do not represent the
differences composing a feature.  The excursion test ignores up
to the $ N_{top} $ largest steps if they occur at the start or end
of the spacing.  In practice $ N_{top} = 8 \text{ or } 10 $ amounts
to ignoring outliers at 3 or 4 standard deviations.  Large
differences in the middle, such as those noted in
Figure~\ref{fig:edi}, do contribute to the pool's distribution and
are not ignored.  \ref{algo:excurht} in Appendix~\ref{app:excur}
specifies the screening, and the test in general.

The excursion test takes as input the draw pool, the draw size or
length of the feature to simulate, and its height.  For peaks we
take as the size the feature width to some fraction of its height,
by default 90\%, rather than between minima.  This avoids including
any flats around either minimum, which would widen the peak and
make it easier to re-create its height.  If the fraction is 50\%
this becomes Full Width at Half Maximum (FWHM).

Even after removing the largest differences the excursion test is
unlikely to duplicate the limited height of a flat over its length.
The test should be judged at the 0.01 level for flats, while the
0.05 level is sufficient for peaks.  It will work for features in
either the low-pass or interval spacing.

\section{Changepoints}
\label{sec:cpt}

Rather than searching for and analyzing features within the spacing, we
can ask more generally where it changes behavior.  Changepoint detection
has a long and rich history, going back to statistical process control
and economic trend analysis in the 1940's.  There are more than 25
maintained packages in R encompassing a variety of approaches:
two-sample comparisons of data statistics; cumulative tracking to
control limits to smooth noise and capture trends; out-of-bounds and
outlier identification based on data models; dynamic partitions of
the data following information criteria metrics; and regression fit
consistency.  Our goal is not to evaluate specific approaches ---
see \cite{niu15} or \cite{truong20} for surveys and
\cite{vandenburg22} for a performance comparison --- but to find a
common set of changepoints among the detectors.

This is a problem in classifier fusion, of combining the results of
different analyses.  The variety of detectors leaves only one fusion
strategy.  The detectors provide a list of points and rarely a
quantitative assessment of their decision.  Instead, they usually
take a significance level or a proxy like the average run length as
a parameter and report just the decision.  We are left with fusing
results based on binary outputs --- a data point is or is not a
changepoint --- which can be done by majority voting \cite{xu92}.

However, the variety of detectors also means that some processing
of individual points list is needed.  They vary considerably in the
quality of their results and implementation.  All produced
unreasonable point lists for some test cases encountered during
development, without an obvious pattern as to when this would
happen.  They can be noisy and inconsistent in placing
changepoints.  Some have problems when the data contains gradual
changes at the edges of flats or within regions of consistent
spacing, especially when the variance changes.  Many do not handle
quantized data, triggering at each new value.  Some are slow and a
few are unstable, in the worst case crashing the R session.

Our voting algorithm, given in Appendix~\ref{app:vote}, tries to be
agnostic about the problem.  We do recommend a minimal set of
detectors: the non-parametric PERT test \rpkg{changepoint.np}
\cite{haynes17}; the Iterative Cumulative Sum of Squares package
\rpkg{ICSS} \cite{inclan94}; and the joint segmentation library
\rpkg{jointseg} \cite{pierre14}.  These performed best on a small
internal comparison of all detectors and offer a complementary set
of analysis strategies.  But they are not required and the voting
algorithm will use any detector available on the system.  It allows
for some uncertainty in the position of changepoints by considering
those that are close to be the same in the vote.

Changepoints do not correspond directly to peaks and the edges of
flats.  They mark the transition between the features.  As a trade-off
to working with the raw spacing, without filtering, we lose the ability
to place modes and anti-modes, unless the changes are very sharp, as
will happen with the first variate of \eqref{eq:trimode}.  Changepoints
complement the feature analysis and indicate the presence of
multi-modality, but they do not allow us to quantify that statement.

\section{Two Examples}
\label{sec:ex}

\begin{figure}
\centering
\begin{minipage}[t]{\textwidth}
\includegraphics[width=\textwidth]{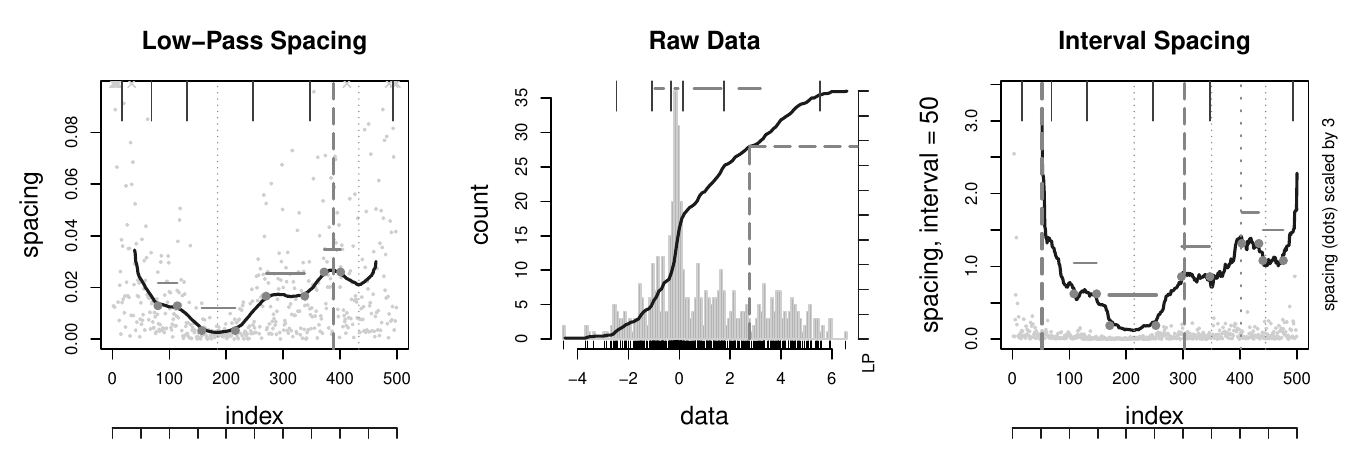}
\caption{\label{fig:triex} Spacing analysis of modality for tri-modal example.
  Compare to the ideal graphs in Figure~\ref{fig:edi}. }
\end{minipage}
\end{figure}

As the first example of the spacing analysis, Figure~\ref{fig:triex}
shows its results for the tri-modal setup \eqref{eq:trimode}.  The
left graph plots the low-pass filtered spacing atop the sample
points, with the same draw used for Figure~\ref{fig:edi}.  The right
graph plots the interval spacing, which is much more uneven although
it still has the same overall shape.  Vertical dashed lines mark peaks
and dotted lines minima. Flats are marked with bars above thick dots at
the endpoints.  If either feature passes testing, it is drawn with a
thick line. The ticks at the top of the graph mark changepoints.  The
center graph is a histogram of the data with the distribution function
drawn on top. The axis to the right marks deciles and matches the axes
under the spacing graphs, to convert index positions to data values.
The annotations for the peaks, flats, and changepoints in this graph
match the low-pass spacing.

The distribution of the actual draw follows the ideal in
Figure~\ref{fig:edi}, although the second and third variates are
less distinct.  The histogram is a noisy version of the distribution
graph, which obscures the anti-mode.  The actual spacing follows the
expected, but the left and right shoulders do not match and the
tails are cut off by the filters.  The different indexing
conventions for the low-pass filter and interval are seen in the
horizontal position of the curve.

The low-pass spacing contains one peak at index 389 (raw value 2.771)
which shifts $ +0.294 $ compared to the peak in the expected spacing.
It passes both the height model and excursion test with $ p = 0.000 $
(rounded to three decimal places).  The interval spacing has three
peaks.  That at index 402 (2.493) matches the low-pass peak, taking
the different indexing of the interval spacing into account.  It does
not pass the excursion or any of the runs tests, however.  Instead,
the peak at index 302 (0.6344) passes the run length test in the
rise from the flat, with $ p = 0.001 $, and the run height
permutation test with $ p = 0.003 $.  This point, and its neighboring
minimum at index 350 (1.4487), are glitches in the rough right
shoulder that are smoothed away by the Kaiser filter.  Had they not
appeared then the peak at 402 would have passed the longest run
test, and marginally the excursion.  The third peak at index 52
($ -2.1574 $) passes the excursion test with $ p = 0.000 $ because of
the large drop to the tight mode.  It too is noise, a single point
of increase at the edge of the data that is not merged in the
detector because the first data point, here a minimum, is anchored
as an extrema and cannot be deleted.

The flats pick out the tight variate, spanning indices 157--216 
($ -0.190 $ -- $ -0.035 $).  It is significant in the interval spacing,
passing the excursion test with $ p = 0.003 $, but not in the low-pass
spacing.  Flats also pick out the shoulders.  The right one in the
interval spacing marginally fails the excursion test in the interval
spacing with $ p = 0.033 $.  A flat appears within the peak in both
spacings, and in the third mode in the interval spacing.  None of
these pass testing.

Changepoints at indices 131 ($ -0.3264 $) and 247 (0.1409) bound the
tight variate, with its sudden change in variance.  One at index
347 (1.7591) lies to the left side of the peak separating the
second and third variates, marking the transition between modes
but not the peak itself. Changepoints often appear in the two
tails of the sample, as happens here, and there is a sixth at
index 69 ($ -1.0762 $) at an outlier in the left shoulder.

\begin{table}
\begin{center}
\caption{\label{tbl:triex} Tri-Modal Feature Stability Over 30 Different Draws}
{\small
\begin{tabular}{llcccc}
$ \quad $ & & \multicolumn{2}{c}{low-pass} &
  \multicolumn{2}{c}{interval} \\
\multicolumn{6}{l}{peaks} \\
 & at anti-mode                & 24 & (24) & 20 & (16) \\
 & elsewhere                   &  0 & (0)  & 18 & (12) \\
\multicolumn{6}{l}{flats} \\
 & in tight variate            & 30 & (0)  & 21 & (17) \\
 & in left shoulder            & 7  & (0)  & 13 & (1)  \\
 & in right shoulder           & 15 & (0)  & 22 & (3)  \\
 & in third variate            & 16 & (0)  &  7 & (0)  \\
 & at peak                     & 17 & (0)  &  7 & (1)  \\
\multicolumn{6}{l}{changepoints} \\
 & both sides of tight draw ~~ & 27 \\
 & near peak                   & 15 \\
 \\
\multicolumn{6}{l}{{\small counts of features passing any test in parentheses}}
 \\
\end{tabular}
}
\end{center}
\end{table}

The spacing does resolve this multi-modal setup.  It locates the
peak between the two main normals, and the edges of the third,
tight draw. It is consistent in the shoulders to the sides of
this variate, enough to form flats, although they do not pass
testing.  The bottom of the right variate is too rounded in the
low-pass spacing by the transition to the tail, however.  But
this draw was naturally chosen to demonstrate these features.
If we generate the sample for 30 different seeds of the random
number generator, not all will show them. Table~\ref{tbl:triex}
counts in how many trials a feature appears, and if they pass
any test.  The peak at the anti-mode is stable, and significant
in the low-pass spacing but only half the time in the interval
spacing. There is also a second significant interval peak in
half the samples. The low-pass spacing has a stable flat in the
tight variate, and less repeatably in the third variate and
even over the anti-mode, but testing rejects all.  The tight
variate has a flat in the interval spacing that is less stable,
but it is accepted when it appears.  Occasional interval flats
in the third variate and at the anti-mode are rejected.  The
right shoulder from the second variate often generates a flat,
the left less often.  None of the shoulder flats pass testing.
No flats cover more than one feature.  Changepoints on both
sides of the tight variate are stable, but the peak is marked
in only half the trials.

\begin{figure}
\centering
\begin{minipage}[t]{\textwidth}
\includegraphics[width=\textwidth]{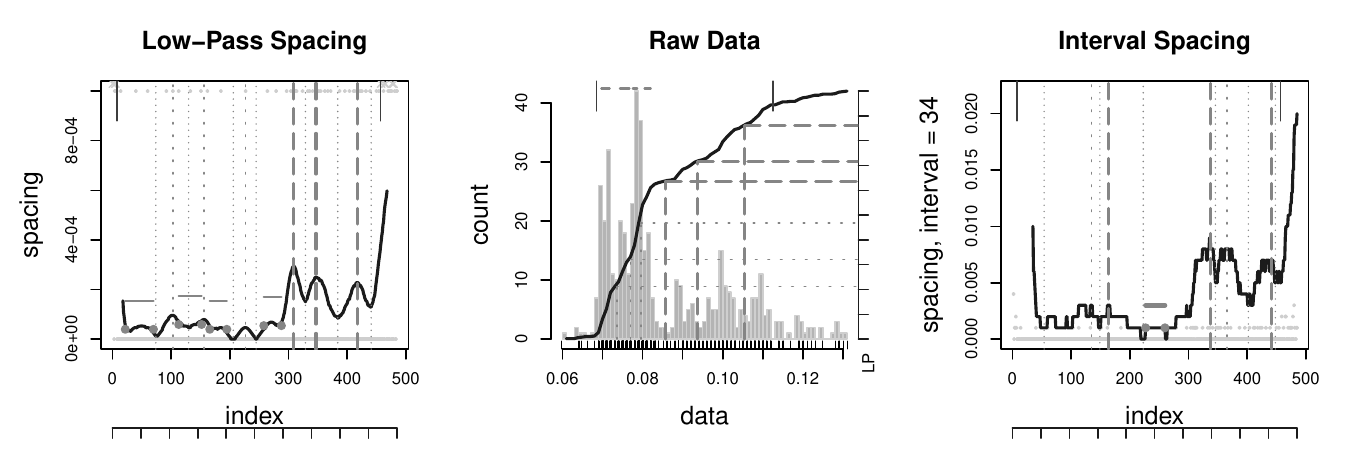}
\caption{\label{fig:stamps} Spacing analysis of modality of Hildago stamp
  issue.}
\end{minipage}
\end{figure}

For the second example we turn to the thickness of the paper of
the 1872 Hildago stamp issue in Mexico, which has been analyzed
for multi-modality.  \cite{izenman88} found seven modes by using
mixture models and three by another method, and \cite{basford92}
confirmed the seven modes.  485 stamps were measured to micron
tolerance, with values ranging from 0.060 to 0.131~mm.  This
strongly quantizes the spacing.  423 points have zero spacing,
54 step by 0.001, 5 by 0.002, and one each by 0.003 and 0.004.
Figure~\ref{fig:stamps} provides the analysis, following the
same format as the tri-modal example.  The limited resolution
is clear not only in the placing of the raw spacing points in
the background, but also in the steps in the interval spacing,
all of size 0.001.  Still, the interval spacing resembles the
low-pass curve, with the same features.  To pick up the smaller
features we use a filter size of $ f_{lp} = 0.07 $ and a
matching interval width.  We must change $ f_{ht} = 0.10 $ for
the peak detector in order to avoid each step forming a peak in
the interval spacing.

The low-pass spacing contains six peaks, with a seventh at the left
side ignored for having too small a drop to its left minimum.  The
peak at index 309 (thickness 0.0858) passes both the excursion test
with $ p = 0.001 $ and height model with $ p = 0.002 $.  At indices
347 (0.0938) and 417 (0.1054) the excursion test passes with
$ p = 0.019 $ and $ 0.045 $ respectively.  The other three peaks at
indices 103 (0.0736), 156 (0.0770), and 227 (0.0795) are not
accepted.

The interval spacing contains five peaks.  Three at thicknesses of
0.0764, 0.0892, and 0.1074 pass the run length test with $ p = 0.005 $,
$ 0.003 $, and $ 0.009 $ respectively.  The second also has a passing
run height test probability of $ p = 0.005 $.  None of the other tests
--- excursion, number of runs, or runs permutation --- comes close to
the passing threshold.  The two peaks that pass no tests lie at
0.0747 and 0.0944.  Notice in the figure how the roughness of the
interval spacing and quantization of the data makes locating the peaks
difficult, with broad uneven tops around indices 337, 366, and 442
and a drop-out between 135 and 164.  Relaxing the $ f_{h,tie} $
parameter would treat these as equal.  The uncertainty complicates
the matching of results between the two spacings.

The interval spacing contains one flat, the low-pass spacing four.
Only the interval flat, extending between 0.0791 and 0.0800, passes
the excursion test with $ p = 0.001 $.  The first low-pass flat,
between 0.0697 and 0.0718, marginally fails the excursion test with
$ p = 0.015 $.  The other flats, between 0.0744 -- 0.0767, 0.0776 --
0.0787, and 0.0805 -- 0.0820, are not significant, although
they seem to match modes in the histogram.

The minima in the low-pass spacing match the location of the modes
reported in the original papers (Table~\ref{tbl:stamps}).  Passing
peaks separate these minima.  Those with thicknesses 0.120 and 0.129
are lost in the final tail to the right of both the low-pass and
interval spacing.  Both spacings have an additional minimum at
0.0754 (index 130) for the bars in the histograms between the major
peaks at 0.07 and 0.08.  The original papers do not include this as
a mode.

\begin{table}
\begin{center}
\caption{\label{tbl:stamps} Modes in the 1972 Hidalgo Stamp Issue}
{\small
\begin{tabular}{lrrrrrrr}
spacing \\
 \quad minima
   & 0.0719 & 0.0790 & 0.0902 & 0.1004 & 0.1097 & & \\
Izenman \\
 \quad method 1
   & 0.0723 & 0.0797 & 0.0905 & 0.1002 & 0.1095 & 0.1208 & 0.1293 \\
 \quad method 2
   & 0.0712 & 0.0786 &        & 0.0989 \\
Basford \\
   & 0.0716 & 0.0792 & 0.0907 & 0.1003 & 0.1096 & 0.1202 & 0.1285
\end{tabular}
}
\end{center}
\end{table}

Changepoints are not useful for this example.  The common points are
placed in the tails, responding to the largest spacings.  A look at
the individual detector results show that seven of twelve respond near
index 300 (0.0830), but the position is not stable within a separation
bound of 10 points.  One detector, the WBS method of the \rpkg{cpss}
package, finds four points at 0.0705, 0.0723, 0.0797, and 0.0892,
which matches the first three modes.

As a third example, the \rpkg{Dimodal} package contains a data set
with the orbital axes of asteroids.  Peaks correspond to the main
Kirkwood gaps and flats indicate families of bodies.
\cite{kreider25c} documents the analysis.

\section{Performance}
\label{sec:perf}

\cite{kreider25b} evaluates how well the features represent and locate
changes in modality and how effective the tests are at judging them.
It updates the key results of the complete analysis found in
\cite{kreider24} with the final implementation of this work, in the
\rpkg{Dimodal} package.  These papers provide the details behind the
claims made in this summary.

\begin{figure}
\centering
\begin{minipage}[t]{\textwidth}
\includegraphics[width=\textwidth]{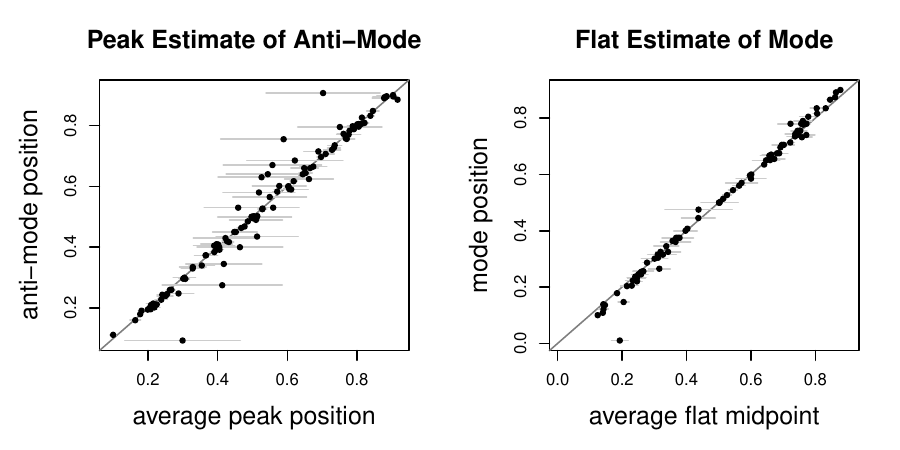}
\caption{\label{fig:featpos} Comparison of low-pass peaks and flat midpoints
  against ideal feature positions of multi-modal examples. }
\end{minipage}
\end{figure}

Flats and peaks track the position of modes and anti-modes, using the
midpoint of the span as the estimator of the mode
(Figure~\ref{fig:featpos}).  The graphs plot the position of the
low-pass features against the ideal for 81 uni- and multi-modal
samples collected from various papers describing modality tests.
The ideal positions come from the numeric integration of \eqref{eq:edi}
for each sample.  The positions are averaged over 400 draws of each
sample. Bars cover one standard deviation of the separation between
the ideal and actual positions.  Both peaks and flat midpoints
follow the 45 degree line that represents an exact match. Several
peaks have long bars representing a large uncertainty in their
position.  The instability can be seen in
\cite[Figure 8]{kreider25b} as extended greyscale counts.  It
occurs in asymmetric setups from the variation in one or two small
draws next to a much larger, stable variate.

Feature positions are correct even when the tests do not pass their
acceptance thresholds, although they become progressively less stable
as the modes become less distinct.  The presence of passing peaks is
enough for a uni-modality test.  Changepoints lie to the side of
modes and anti-modes, especially the latter, and do not mark them
directly.  They often flag only one side of a peak and respond best
to very sharp changes in modality, as we have seen in the tri-modal
example. Changepoints often occur in the initial or final tails,
responding to the large changes in the arms of the `U'.

The interval spacing is rough, which means it will contain more peaks.
Testing reduces the difference to the low-pass spacing results, so
that accepted feature counts are similar; half of the detected
low-pass peaks pass, while a third in the interval spacing do.  The
positional accuracy in the interval spacing is worse. The agreement
of a peak's location in the two spacings is better for detected
peaks than after testing, helped by the greater count in the
interval spacing but also implying that there is some randomness in
the position due to the roughness.  The peak detector parameters
may need tweaking to compensate.  Flats in the interval spacing
match those in the low-pass, in endpoint location and stability and
overall count and acceptance rate.  They mark modes.

Tests on the low-pass spacing do resolve modality with a fairly sharp
transition.  When the rise in the expected spacing at the anti-mode
between two normals is 0.003, due to any mismatch in the distribution
parameters, then the height model or excursion test will accept the
peak.  The height model decision for or against uni-modality behaves
similarly to the excess mass test \cite{ameijeiras16}, although it
can resolve smaller differences. The excursion test behaves
similarly, but if the test threshold is changed to 0.01 to sharpen
the decision and reduce false positives it needs somewhat larger
peaks, corresponding to more distinct modes, and resembles the dip
test \cite{hartigan85}.  However, the dip and critical bandwidth
\cite{silverman81} tests are calibrated for uniform variates, and
both spacing tests accept too many peaks from such draws.  They, and
the excess mass test, judge uni-modal normal draws correctly.

The interval spacing tests, both runs and excursions, are less
sensitive to the data's modality.  Variates must be much more
distinct before they can be resolved, needing rises of 0.014, and
the test probabilities have a broad transition as the peak height
changes, so the judgement for or against uni-modality is not crisp.
The longest run and runs count tests are equivalent.  They respond
very slowly to differences between modes. A passing result is
strong evidence for multi-modality, but other tests will likely
pass first.  The run height permutation test is more sensitive
than the excursion test but is biased towards accepting marginal
peaks, giving a false positive rate of 10--20\% even when the
variates cannot be separated.

There is little correlation between the tests.  They can be taken as
an ensemble, used in aggregate.  Any passing result indicates
multi-modality.

\section{Conclusion}
\label{sec:conc}

We have developed and investigated six approaches using spacing to
detect and analyze modality.  They are complementary.  The new tests
are parametric models for peaks and flats after low-pass filtering,
which amounts to working with kernel density estimates, and a Markov
chain-based model of the longest run in the signed difference of the
interval spacing.  Other tests on the signed difference include the
Kaplansky-Riordan statistic on the number of runs, and a comparison
to the reconstructed height of their permutations.  Also extending
known techniques, the excursion test is a bootstrap estimate of a
feature's likelihood using the difference of the filtered spacing
as a pool for its reconstruction.  A voting algorithm combines
existing changepoint detectors to find switches in the spacing's
behavior.  These tests check can not only if data is multi-modal,
they can also locate the modes and transitions between them.
Smoothing the spacing allows us to do this on discrete or heavily
quantized data by responding to the density of changes.

Some open questions about the tests remain.  How does changing detector
parameters affect their performance?  We have noted that this might
be necessary when working with the interval spacing, but do not know
what effects such changes may have. They may change the critical
values for the models, or perhaps less likely, may bias the excursion
tests.  How should one handle very large data sets, on the order of
tens or hundreds of thousands of points?  The spacing will either be
strongly quantized if the data resolution is poor, or it will be
naturally smooth as the limited range of values is divided over such
a large set.  The first condition may require too much smoothing to
cover the steps.  The second may require tighter detector parameters
to handle small changes or setting filter windows very small which
may push the models into invalid regimes.  Significant effort has
gone into optimizing the detectors and tests for these cases, which
make the data-driven tests more useful.  Changepoint analysis will
be at best difficult because many of the libraries do not scale.
The positional analysis has focussed on peaks and anti-modes; can
we process the flats to estimate the mode, similar to Venter's
estimate, as Figure~\ref{fig:featpos} implies may be the case?  Or,
as we saw in the stamps example, the minima of the spacing seemed
to align well with the modes; does this hold in general and is it
stable, particularly in the interval spacing?  Finally, can we
combine this ensemble of tests into one overall result?  We have
probabilities from each individual result, so other fusion
strategies than the majority vote for the changepoint detectors
are possible.

We can also begin to set limitations on this approach.  The models
degrade at the edge of their parameter space, especially for small
data or filters.  The peak height model becomes inaccurate with
$ n \le 70 $ or $ n \ge 500 $ points, or for fractional kernel
sizes $ f_{lp} \le 0.075 $ or $ f_{lp} > 0.30 $.  The flat length
model degrades if $ n \le 60 $ or $ f_{lp} \le 0.075 $.  We know
that initial and final tails in the data produce large increases
in the spacing, and these can and do hide modes at the edges.
This is not just a problem in samples designed to stress a
modality test, as seen in the literature samples, but also
appears in real data, such as the upper two modes in the stamps
data.  Excluding points partially covered by the low-pass filter
hides the problem, and the interval spacing does better at the
edges.  If the data has limited resolution, or in the worst case
is discrete, then the density of step changes places limits on
the minimum filter size and what can be detected.  The
changepoint detectors do not handle these conditions, triggering
at each value change, and the interval spacing tests do not do
much better.  Another restriction is the appearance of flats.
Unless they are created by very strong scale changes, as in the
tri-modal example, consistent spacing does not often produce a
clean, stable filtered signal.  The bottoms round off and slow
transitions at the edge of the flat erode the length. This may
be the classic trade-off between filter size and bandwidth. It
seems to happen in the stamp example and in the rightmost mode
of the tri-modal setup.  Finally, by using the spacing we are
limited to one-dimensional data, as there is no way to rank points
in a higher dimensional space.  We cannot attack the general
clustering problem this way.

\section{Software and Supplementary Material}
\label{sec:sw}

The R package \rpkg{Dimodal} provides the reference implementation
of the feature detectors and tests.  It includes the peak height and
flat length models, runs tests, and excursion evaluation of peaks and
flats.  The package is designed to work with whatever changepoint
detectors are available on the system, without adding a large set of
external dependencies, and provides an additional detector that
inverts the interval spacing test of \cite{rufibach10} to find level
sections in the data.  The package includes a tool similar to the
mode tree \cite{minnotte92} or SiZer graph \cite{chaudhuri00} to help
choose the size of the filter or interval width.  The supplementary
material includes scripts generating the figures and data for this
paper, as well as the literature sample analysis.  Comments within
the main script show how a full  analysis of the stamps data would
be done.

\section{Disclosure of Interest and Funding Statements}
There are no competing interests to declare.  No funding was received
for this work.

\bibliographystyle{tfcad}
\bibliography{dmodal}

\begin{appendix}

\section{Detector Algorithms}
\label{app:detect}

The runs detector in \ref{algo:runs} builds two arrays, one holding
the length of a run at the index it starts and the other a count of
values skipped within the run if the language supports invalid markers,
such as NaN floating point values or NA.  Steps~2, 3, and 4 seed the
initial skip count, then step~5 registers runs within the remaining
data using a second tracking index $ j $.  It counts skipped values
along the way, and breaks the run when values do not match, either
exactly if the data is discrete or according to the relative
matching criterion if not.  Small differences on the order of the
machine tolerance are always treated as equal, even if $ f_{eps} $
is zero.  This may occur when reading floating point data, for
example $ (0.3 - 0.2) - (0.2 - 0.1) $ leaves a residual on the order
of $ 10^{-17} $.

In the pseudo-code, $ \sizeof{x} $ when applied to an array or set
equals its length or size.  Array indexing is one based unless
noted, and ranges of indices are marked with dots.  Curly
brackets form sets and a vertical bar within generates those
values on the left that satisfy the conditions to the right.
A backslash applied to sets removes the second value.

\func{relht} is a utility function that calculates the ratio of
the relative difference between two points to their average,
\begin{equation} \label{eq:relht}
\text{relht}(x, i, j) =
  \lvert x[j] - x[i] \rvert / ((\lvert x[j] \rvert + \lvert x[i] \rvert) / 2)
\end{equation}
The absolute value in the denominator handles points on different
sides of zero.  The value is less useful near zero or if the signal
is noisy so that the difference is much larger than the average.
Because the spacing is always positive the first concern does not
apply to this application.  \ref{algo:runs} also uses functions
\func{is.valid} to judge if a value should be skipped, and
\func{is.discrete} to judge if the data has separate values that
compare through equality.

\begin{center}
\begin{algo}{algo:runs}
\alghdr{{\bf rundet}: Runs Detector} \\
\algrow[Input:]{ signal $ x $, relative matching fraction $ f_{eps} $ }
\algrow[1:]{ let length $ l[1 \ldots \sizeof{x}] \leftarrow 0 $,
                 skip count $ s[1 \ldots \sizeof{x}] \leftarrow 0 $ }
\algrow[2:]{ let $ i \leftarrow 1 $ }
\algrow[3:]{ while $( i < \sizeof{x} )$ AND NOT $ \text{is.valid}( x[i] )$ }
\algrowi[3a:]{ $ i \leftarrow i + 1 $ }
\algrow[4:]{ $ s[1] \leftarrow i $ }
\algrow[5:]{ while $ i \leq \sizeof{x} $ }
\algrowi[5a:]{ let $ j \leftarrow i + 1 $ }
\algrowi[5b:]{ while $ j \leq \sizeof{x} $ }
\algrowii{ if NOT $ \text{is.valid}( x[j] ) $ }
\algrowiii[5b1a:]{ $ s[i] \leftarrow s[i] + 1 $ }
\algrowii{ else if $ \text{is.discrete}(x) $ }
\algrowiii{ if $ x[i] \ne x[j] $ }
\algrowiv[5b1b:]{break inner while}
\algrowii{ else }
\algrowiii{ if $ \text{relht}(x, i, j) \geq f_{eps} $ OR
                   $ \lvert x[j] - x[i] \rvert \geq \text{MACHINETOL} $}
\algrowiv[5b1c:]{break inner while}
\algrowii[5b2:]{ $ j \leftarrow j + 1 $ }
\algrowi[5c:]{ $ l[i] \leftarrow j - i - s[i] $ }
\algrowi[5d:]{ $ i \leftarrow j $ }
\algrow[6:]{ return $ l $, $ s $ }
\end{algo}
\end{center}

A run length encoding of the data contains the non-zero elements of
the returned array $ l $, plus the data values at the corresponding
indices.  To traverse the returned arrays, start at index $ i = 1 $
if $ l[1] \ne 0 $ or at $ 1 + s[1] $, then increment by $ l[i] + s[i] $.
The \rpkg{Dimodal} utility function \func{runs.as.rle} does this
conversion, returning two arrays with the non-zero lengths and
corresponding values.

The peak detector screens all potential local maxima, eliminating
those that are separated by shallow minima from a neighboring, larger
peak.  Shallow here means that the height of the peak above the
minimum is less than a fraction $ f_{ht} $ of the total signal range,
or than a fraction $ f_{relht} $ of the average level of the peak and
minimum, using \eqref{eq:relht}.  The signal is passed first through
the runs detector with epsilon $ f_{h,tie} $ to treat broad but rough
maxima as a point; in practice this rarely happens but a relaxed value
could be useful when processing the interval spacing.

The algorithm in \ref{algo:peaks} includes an in-line version of the
runs detector in steps~2 and 3, but the implementation calls
\func{rundet} and works on its result.  This compresses runs to a
single point, which the tests in steps~4 and 5 assume is the case;
they do not handle tied values.  Instead, in the pseudo-code we use
the $ pos $ array to track the collapse and remove the tied points
in step~3c.  The maxima are processed one by one in increasing height
order in step~7b, using the \func{heights} function in
\ref{algo:heights} that determines the smallest offset to the
left/previous or right/next minimum, storing which in the $ hID $
identifier.  \func{heights} also calculates the relative height
and to which minima it is tied.  Keeping the absolute and relative
heights separate allows for one or both minima to fail in step~7d,
with the absolute requirement taking precedence.  If the peak is too
small, it is removed from the maxima set along with the appropriate
minima.  This will change the neighbors for the next pass through
the loop, and therefore their height. Step~8 generates the final
height information and steps~9 and 10 return the maxima and minima.
The heights may be scaled at this time by the signal's standard
deviation for the model, but not for the excursion test.

\begin{center}
\begin{algo}{algo:peaks}
\alghdr{{\bf peakdet}: Local Maxima Detector} \\
\algrow[Input:]{ signal $ x $,  peak height requirements $ f_{ht} $,
                   $ f_{relht} $, matching $ f_{h,tie} $ }
\algrow[1:]{ let signal range $ x_{h} \leftarrow \max(x) - \min(x) $ }
\algrow[2:]{ let peak position $ pos[1 \ldots \sizeof{x}]
               \leftarrow 1 \ldots \sizeof{x} $ }
\algrow[3:]{ for $ j_{st} $ in $ {1 \ldots} \sizeof{x} $ }
\algrowi[3a:]{ $ j \leftarrow j_{st} $ }
\algrowi[3b:]{ while $ j \leq (\sizeof{x} - 1) $ AND }
\algrowiii[]{ ( $ \text{relht}(x, j, j+1) \leq f_{h,tie} $  OR }
\algrowiii[]{ ~ $ \sizeof{x[j] - x[i]} < \text{MACHINETOL} $ ) }
\algrowii[3b1:]{ $ j \leftarrow j + 1 $ }
\algrowi{ if $ j_{st} < j $ }
\algrowii[3c1:]{ $ pos[j_{st}] \leftarrow (j_{st} + j) / 2 $ }
\algrowii[3c2:]{ delete $ x[(j_{st}+1) \ldots j] $,
                        $ pos[(j_{st}+1) \ldots j] $ }
\algrow[4:]{ let $ \sets{I}_{min} \leftarrow \{ i ~\lvert~
                   x[i] < x[i-1] \text{ AND } x[i] < x[i+1] \} $ }
\algrow[5:]{ let $ \sets{I}_{max} \leftarrow \{ i ~\lvert~
                   x[i] > x[i-1] \text{ AND } x[i] > x[i+1] \} $ }
\algrow[6:]{ let $ \sets{I}_{todo} \leftarrow \sets{I}_{max} $ }
\algrow[7:]{ while $ \sets{I}_{todo} \neq \varnothing $ }
\algrowi[7a:]{ let $ h, hID, relht, relhtID \leftarrow
                   \text{heights}(x, \sets{I}_{todo}, \sets{I}_{min}) $ }
\algrowi[7b:]{ let $ i \leftarrow
                   \arg\min_{k \in 1 \ldots \sizeof{\sets{I}_{todo}}} h[k] $ }
\algrowi[7c:]{ $ \sets{I}_{todo} \leftarrow
                     \sets{I}_{todo} ~\backslash~ \sets{I}_{todo}[i] $ }
\algrowi{ if $ h[i] < x_{h} f_{ht} $ }
\algrowii[7d1:]{ $ \sets{I}_{min} \leftarrow
                     \sets{I}_{min} ~\backslash~ hID[i] $ }
\algrowii[7d2:]{ $ \sets{I}_{max} \leftarrow
                     \sets{I}_{max} ~\backslash~ \sets{I}_{todo}[i] $ }
\algrowi{ else if $ relht[i] < f_{relht} $ }
\algrowii[7d3:]{ $ \sets{I}_{min} \leftarrow
                     \sets{I}_{min} ~\backslash~ relhtID[i] $ }
\algrowii[7d4:]{ $ \sets{I}_{max} \leftarrow
                     \sets{I}_{max} ~\backslash~ \sets{I}_{todo}[i] $ }
\algrow[8:]{ let $ h, hID, relht, relhtID \leftarrow
                   \text{heights}(x, \sets{I}_{max}, \sets{I}_{min}) $ }
\algrow[9:]{ report maxima in $ \sets{I}_{max} $: }
\algrowi{ position $ pos[\sets{I}_{max}] $, value $ x[\sets{I}_{max}] $,
          height $ h[\sets{I}_{max}] $ }
\algrow[10:]{ report minima in $ \sets{I}_{min} $: }
\algrowi{ position $ pos[\sets{I}_{min}] $, value $ x[\sets{I}_{min}] $,
          height $ h[\sets{I}_{min}] $ }
\end{algo}
\end{center}

In the implementation $ \sets{I}_{min} $ and $ \sets{I}_{max} $ are
interwoven in linked lists so that determining neighbors and deleting
pairs is easy.  Heights are updated incrementally after setting up,
and the peaks are maintained in a minheap by height so the smallest
is immediately available and updates are efficient.  The detector
always keeps the first and last data points as extrema even if the
height is small.  This requires careful handling in the implementation
to avoid edge cases.

\begin{center}
\begin{algo}{algo:heights}
\alghdr{{\bf heights}: Determine peak height to neighboring minima} \\
\algrow[Input:]{ signal $ x $, indices of local peaks $ \sets{I}_{max} $ and
                 minima $ \sets{I}_{min} $ }
\algrow[1:]{ create array height $ h[1 \ldots \sizeof{\sets{I}_{max}}] $,
                index $ hID[1 \ldots \sizeof{\sets{I}_{max}}] $ }
\algrow[2:]{ create array $ relht[1 \ldots \sizeof{\sets{I}_{max}}] $,
                index $ relhtID[1 \ldots \sizeof{\sets{I}_{max}}] $ }
\algrow[3:]{ for $ i $ in $ 1 \ldots \sizeof{\sets{I}_{max}} $ }
\algrowi[3a:]{
  let left minima $ l \leftarrow
           \arg\max_{l \in \sets{I}_{min}} l < \sets{I}_{max}[i] $ }
\algrowi[3b:]{
  let right minima $ r \leftarrow
           \arg\min_{r \in \sets{I}_{min}} r > \sets{I}_{max}[i] $ }
\algrowi[3c:]{ let $ h_{L} \leftarrow x[\sets{I}_{max}[i]] - x[l] $ }
\algrowi[3d:]{ let $ h_{R} \leftarrow x[\sets{I}_{max}[i]] - x[r] $ }
\algrowi{ if $ h_{L} < h_{R} $ }
\algrowii[3e1:]{ $ h[i] \leftarrow h_{L} $ ;
                 $ hID[i] \leftarrow l $ }
\algrowi{ else }
\algrowii[3e2:]{ $ h[i] \leftarrow h_{R} $ ;
                 $ hID[i] \leftarrow r $ }
\algrowi[3f:]{
  let $ relht_{L} \leftarrow \text{relht}(x, \sets{I}_{max}[i], l) $ }
\algrowi[3g:]{
  let $ relht_{R} \leftarrow \text{relht}(x, \sets{I}_{max}[i], r) $ }
\algrowi{ if $ relht_{L} < relht_{R} $ }
\algrowii[3h1:]{ $ relht[i] \leftarrow relht_{L} $ ;
                 $ relhtID[i] \leftarrow l $ }
\algrowi{ else }
\algrowii[3h2:]{ $ relht[i] \leftarrow relht_{R} $ ;
                 $ relhtID[i] \leftarrow r $ }
\algrow[4:]{ return $ h $, $ hID $, $ relht $, $ relhtID $ }
\end{algo}
\end{center}

The flat detector in \ref{algo:flats} makes two passes through the
data.  In the first it scans left and right from each source point
using the function \func{extendflat} in \ref{algo:extflat}.  The
scan moves outward in step~5 from the point until it encounters
either $ noutlier $ points that fall outside the ripple
specification, which is centered about the source point, or it
reaches the end of the data.  In step~6 it backs off the endpoint
until it reaches a valid point; the flat cannot end on an outlier
and the source point $ i $ is sure to be valid.  The detector then
determines the valid flats in the second phase.  Working from
longest to shortest, it counts the number of points covered by the
flat that are not part of a longer one.  These open segments may
be disjoint.  If the count is above the minimum length requirement
the detector reports the original information about the flat, and
not just the uncovered portion.  This means flats can overlap, as
long as the extension meets the requirement.  The phase stops in
step~8a when the remaining flats are shorter than allowed.

The function \func{order} returns the indices to sort $ x $ in an
increasing or decreasing direction.  \func{sort} does the actual
sorting.

Both the scan in step~5 and count in step~8c check each point within
a flat.  If flats make up most of the data, which becomes true as the
data size increases and spacing naturally becomes smaller and smoother,
this detector requires $ O({\sizeof{x}}^{2}) $ time.

\begin{center}
\begin{algo}{algo:flats}
\alghdr{{\bf flatdet}: Flats Detector} \\
\algrow[Input:]{ signal $ x $, ripple limit $ f_{ripple} $,
                 absolute minimum length $ L_{min} $, }
\algrow{ relative minimum length $ f_{minlen} $,
         outliers allowed $ noutlier $ }
\algrow[1:]{ let signal range $ x_{h} \leftarrow \max(x) - \min(x) $ }
\algrow[2:]{ let ripple specification
  $ \Delta \leftarrow f_{ripple} x_{h} / 2 $ }
\algrow[3:]{ let length requirement
  $ L \leftarrow \max(f_{minlen} \sizeof{x},\, L_{abs}) $ }
\algrow[4:]{ create endpoint arrays $ st[1 \ldots \sizeof{x}] $,
               $ end[1 \ldots \sizeof{x}] $ }
\algrow[5:]{ create seed index array $ src[1 \ldots \sizeof{x}] $,
               length $ len[1 \ldots \sizeof{x}] $ }
\algrow[6:]{ for $ i $ in $ 1 \ldots \sizeof{x} $ }
\algrowi[6a:]{ $ src[i] \leftarrow i $ }
\algrowi[6b:]{ $ st[i] \leftarrow
                  \text{extendflat}(x, i, 1, \Delta, noutlier) $ }
\algrowi[6c:]{ $ end[i] \leftarrow
                  \text{extendflat}(x, i, \sizeof{x}, \Delta, noutlier) $ }
\algrowi[6d:]{ $ len[i] \leftarrow end[i] - st[i] + 1 $ }
\algrow[7:]{ create array $ open[1 \ldots \sizeof{x}] \leftarrow 1 $ }
\algrow[8:]{ for $ i $ in $ \text{order}(len, \text{DECREASING}) $ }
\algrowi{ if $ len[i] < L $ then }
\algrowii[8a1:]{ break for }
\algrowi[8b:]{ let $ nopen \leftarrow 0 $ }
\algrowi[8c:]{ for $ j $ in $ st[i] \ldots end[i] $ }
\algrowii[8c1:]{ $ nopen \leftarrow nopen + open[j] $ }
\algrowii[8c2:]{ $ open[j] \leftarrow 0 $ }
\algrowi{ if $ L \leq nopen $ }
\algrowii[8d1:]{ report flat source $ src[i] $, extent $ st[i] $ to $ end[i] $,
                    length $ len[i] $ }
\end{algo}
\end{center}

\begin{center}
\begin{algo}{algo:extflat}
\alghdr{{\bf extendflat}: Determine endpoint of flat}
\algrow[Input:]{ signal $ x $, starting index $ i $, stopping index $ s $, }
\algrow{ one-sided ripple $ \Delta $, maximum outlier count $ noutlier $ }
\algrow{ if $ i = s $ then }
\algrowi[1a:]{ return $ i $ }
\algrow{ if $ i < s $ then }
\algrowi[2a:]{ let $ dir \leftarrow +1 $ }
\algrow{ else }
\algrowi[2b:]{ let $ dir \leftarrow -1 $ }
\algrow[3:]{ let $ j \leftarrow i $ }
\algrow[4:]{ while $ j \ne s $ }
\algrowi{ if $ x[j] < x[i] - \Delta $ OR $ x[j] > x[i] + \Delta $ then }
\algrowii[4a1:]{ $ noutlier \leftarrow noutlier - 1 $ }
\algrowii{ if $ noutlier < 0 $ then }
\algrowiii[4a2a:]{ break while }
\algrowi[4b:]{ $ j \leftarrow j + dir $ }
\algrow[5:]{ while $ j \ne i $ AND
              ( $ x[j] < x[i] - \Delta $ OR $ x[j] > x[i] + \Delta $ ) }
\algrowi[5a:]{ $ j \leftarrow j - dir $ }
\algrow[6:]{ return $ j $ }
\end{algo}
\end{center}

\section{Feature Reconstruction Tests}
\label{app:excur}

The permutation test in \ref{algo:permht} begins by converting the
feature between the given bounds into a set of runs in the signed
difference.  The implementation steps through the \func{rundet}
result from \ref{algo:runs}, but conceptually this is reduced to
two lists with the run lengths $ l $ and directions of change $ v $
with values $ -1 $, $ +1 $, or $ 0 $.  The height in step~1 assumes
the feature is everywhere above the start or finish value, which is
true for this application; otherwise the definition in step~4c
would be appropriate.  Then, sampling the possible permutations
$ N_{perm} $ times, the test calls the \func{altperm} function in
\ref{algo:altperm} to generate a permutation, returning the signed
lengths in permuted order.  It reconstructs the feature in step~4b
using a cumulative sum and determines the height in 4c.  In the
actual implementation the permutations are replaced by an
exhaustive set if the feature is small enough, then screened for
adjacent same symbols.  The quantile returned in step~5 is the
fraction of permutations generating a large enough height.  If
there are ties in the permuted height the quantile is taken at
the midpoint of the tier, a mid-distribution correction.

\begin{center}
\begin{algo}{algo:permht}
\alghdr{{\bf permht}: Run Height Permutation Test} \\
\algrow[Input:]{ signal $ x $, feature index bounds $ j $, $ k $ inclusive,
                   trial count $ N_{perm} $
}
\algrow[1:]{ let feature height
               $ H \leftarrow \max(x[j \ldots k]) - \min(x[j \ldots k]) $ }
\algrow[2:]{ let signed signal $ S \leftarrow \text{sign}(x[i] - x[i-1]) $
               for $ i $ in $ 2 \ldots |x| $ }
\algrow[3:]{ let runs $ R \leftarrow \text{rundet}(S) $
               giving lengths $ l $, values $ v $ }
\algrow[4:]{ repeat $ N_{perm} $ times }
\algrowi[4a:]{ $ r_{perm} \leftarrow \text{altperm}(R) $ }
\algrowi[4b:]{ for $ i' $ in $ 1 \ldots |R| $ }
\algrowii[4b1:]{
  $ x_{perm}[i'] = \sum_{i=1}^{i'} r_{perm}[i] $ }
\algrowi[4c:]{
  $ h_{perm} \leftarrow \max(x_{perm}) - \min(x_{perm}[1],\, x_{perm}[\sizeof{x}]) $ }
\algrow[5:]{
  return $ q_{feat} \leftarrow \left( \#\{h_{perm} < H\} +
                                      \#\{hperm = H\} / 2 \right) / N_{perm} $ }
\end{algo}
\end{center}

The alternating permutation algorithm works for any number of symbols.
In the first phase it alternates them.  If there are only two, the
majority symbol goes in the odd indices of the permutation; ties are
chosen randomly.  There is an assumption here that the symbol
populations differ at most by one, or else alternating symbols is
impossible, and this is satisfied by the application. If there are
more than two symbols, one not matching the previous symbol is picked
randomly in step~4c2 according to the remaining population frequency
until one symbol attains the majority.  At this point the majority
symbol must alternate with any of the others, shuffled freely amongst
themselves in step~4b.  Step~4b1 builds an array of the non-majority
symbol identifiers. The second phase in steps~6 and 7 shuffles the
lengths within each symbol and inserts them, multiplied by the sign,
in the final permutation according to the indices from the first
phase.

The external functions used in \ref{algo:altperm} are \func{U}
returning a uniform variate between two values, \func{shuffle}
that scrambles the order of an array, \func{repeat} that creates
an array with $ N_{t} $ copies of a value, \func{unique} that
returns the unique symbols in the set, and \func{pick} that
chooses a random symbol according to the populations $ N_{u} $
that also does not match the old symbol.  The implementation
tracks $ N_{u} $ incrementally rather than re-counting in each pass
per step~5a.  Note that the symbol set $ \sets{S} $ contains an
element for each run and is not just the unique symbols, which are
extracted in step~2.  In step~5c1 the initial point $ i = 1 $ is
handled by adding an invalid dummy symbol to the start of $ P_{s} $
so that there is no match and the choice is made freely.

\begin{center}
\begin{algo}{algo:altperm}
\alghdr{{\bf altperm}: Generation of Alternating Permutations} \\
\algrow[Input:]{ symbol set $ \sets{S} $, lengths $ l[1 \ldots n] $
                 belonging to symbol values $ v[1 \ldots n] $ }
\algrow[1:]{ create symbol permutation array $ P_{s}[0 \ldots n] $
                 with $ P_{s}[0] \leftarrow \varnothing $ }
\algrow[2:]{ let unique symbol set
                 $ \sets{U} \leftarrow \text{unique}(\sets{S}) $ }
\algrow[3:]{ let symbol count $ N_{u} \leftarrow
                 \#\{ \sets{S} = u \} \text{ for } u \in \sets{U} $ }
\algrow{ if $ \sizeof{\sets{U}} = 2 $ then }
\algrowi{ if $ N_{1} > N_{2} $ OR
             $( N_{1} = N_{2} \text{ AND } U(0,1) < 0.5 )$ then }
\algrowii[4a1:]{ $ j \leftarrow 1 $ ; $ k \leftarrow 2 $ }
\algrowi{else}
\algrowii[4a2:]{ $ j \leftarrow 2 $ ; $ k \leftarrow 1 $ }
\algrowi[4b:]{ $ P_{s}[i=1 \ldots n \text{, } i \text{ odd} ] \leftarrow j $ }
\algrowi[4c:]{ $ P_{s}[i=2 \ldots n \text{, } i \text{ even} ] \leftarrow k $ }
\algrow{ else }
\algrowi[5a:]{ for $ j $ in $ 1 \ldots n $ }
\algrowii{ if any $ N_{u} > (n - j + 1) / 2 $ then }
\algrowiii[5a1a:]{ let minority $ M = \{ \text{repeat}(t, N_{t}) ~\lvert~
                                         t \in \sets{U}, t \ne u \} $ }
\algrowiii[5a1b:]{ $ P_{s}[i=j \ldots n \text{, } i-j \text{ even} ]
                    \leftarrow u $ }
\algrowiii[5a1c:]{ $ P_{s}[i=j \ldots n \text{, } i-j \text{ odd} ]
                    \leftarrow \text{shuffle}( M )$ }
\algrowiii[5a1d:]{ break for }
\algrowii{ else }
\algrowiii[5a2a:]{ $ P_{s}[j] \leftarrow \text{ pick} ( N_{u}, P_{s}[j-1] ) $ }
\algrowiii[5a2b:]{ $ \sets{S} \leftarrow \sets{S} ~\backslash~ P_{s}[j] $ }
\algrowii[5a3:]{ $ N_{u}
                   \leftarrow \#\{\sets{S} = u\} \text{ for } u \in \sets{U} $ }
\algrow[6:]{ create signed length permutation $ P_{lv}[1 \ldots n] $ }
\algrow[7:]{ for $ u \in \sets{U} $ }
\algrowi[7a:]{ let matching indices $ uID \leftarrow \{ i ~\lvert~
                   i = 1 \ldots \sizeof{\sets{S}} \text{ AND } v[i] = u \} $ }
\algrowi[7b:]{ $ P_{lv}[uID] \leftarrow \text{shuffle}(l[uID] \cdot v[uID]) $ }
\algrow[8:]{ return $ P_{lv} $ }
\end{algo}
\end{center}

The excursion height test, \ref{algo:excurht}, has a similar
framework as the permutation test.  The preparation of the draw pool
in steps~1 through 5 involves the difference of the signal and not
its sign.  Step~4 removes up to the largest $ N_{top} / 2 $
differences at the start of the signal, and step~5 does so at the
end.  This leaves large differences within the signal but not in the
strong tails.  It assumes the tails are roughly the same, or an
unequal split could be used if the spacing is strongly asymmetric.
An alternative might define an outlier parameter as a number of
standard deviations above the signal's minimum and remove points at
the start or end until the signal comes within the bound.  In
practice the default values for $ N_{top} $ amount to this
requirement with the cut-off at 3--4 sigma.  The sampling in
step~7a is with replacement, with \func{Uint} generating an integer
drawn uniformly between two bounds, inclusive. The upper/lower tail
flag $ isPeak $ not only controls the definition of the feature
height in step~7 but also inverts the count for the quantile in
step~8.

\begin{center}
\begin{algo}{algo:excurht}
\alghdr{{\bf excurht}: Height Excursion Test} \\
\algrow[Input:]{ signal $ x $, feature index bounds $ j $ to $ k $ inclusive,
                   trial count $ N_{excur} $, }
\algrow{         boolean $ isPeak $, drop count $ N_{top} $ }
\algrow[1:]{ let feature height
               $ H \leftarrow \max(x[j \ldots k]) - \min(x[j \ldots k]) $ }
\algrow[2:]{ let difference signal $ D \leftarrow x[i] - x[i-1] $
                for $ i $ in $ 2 \ldots \sizeof{x} $ }
\algrow[3:]{ let $ o \leftarrow \text{order}(D, \text{DECREASING}) $ }
\algrow[4:]{ $ D \leftarrow D ~ \backslash ~ D[o ~\lvert~ o \le N_{top} / 2] $ }
\algrow[5:]{ $ D \leftarrow D ~ \backslash ~
                          D[o ~\lvert~ o \ge \sizeof{x} - (N_{top} / 2) + 1] $ }
\algrow[6:]{ create array $ x_{excur}[1 \ldots (k-j+1)] $,
                    sample $ s[1 \ldots (k-j+1)] $ }
\algrow[]{ repeat $ N_{excur} $ times }
\algrowi[7a:]{ for $ i $ in $ 1 \ldots k-j+1 $ }
\algrowii[7a1:]{ $ s[i] \leftarrow D[Uint(1, \sizeof{D})] $ }
\algrowi[7b:]{ for $ i $ in $ 1 \ldots k - j + 1 $ }
\algrowii[7b1]{ $ x_{excur}[i] \leftarrow \sum_{i'=1}^{i} s[i'] $ }
\algrowi[]{ if $ isPeak $ }
\algrowii[7c1:]{ $ h_{excur} \leftarrow
                     \max(x_{excur}) - \min(x_{excur}[1], x_{excur}[k-j+1]) $ }
\algrowi[]{ else }
\algrowii[7c2:]{ $ h_{excur} \leftarrow \max(x_{excur}) - \min(x_{excur}) $ }
\algrow[]{ if $ isPeak $ }
\algrowi[8a:]{ return $ p_{feat} \leftarrow
            \left( \#\{h_{excur} < H\} + \#\{h_{excur} = H\}/2 \right) / N_{excur} $ }
\algrow{ else }
\algrowi[8b:]{ return $ p_{feat} \leftarrow
            \left( \#\{h_{excur} > H\} + \#\{h_{excur} = H\}/2 \right) / N_{excur} $ }
\end{algo}
\end{center}

\section{Changepoint Voting}
\label{app:vote}

The majority voting scheme of \ref{algo:cpt} combines changepoints
identified by some collection $ \sets{C} $ of independent detectors,
each with one or more analysis methods $ M $.  It ignores detectors
that generate too few or too many points, either as a fraction $ f_{pt} $
of the data or compared to their peers according to the quantile range
pair $ q_{vote} $.  Points are considered the same if they are close,
either as a fraction $ f_{\Delta} $ of the data length or in absolute
terms $ \Delta_{max} $; the function \func{remove\_nearby} in
\ref{algo:rmnear} does the merging.  The intra-library merge distance
$ \Delta_{lib} $ is typically a little smaller.

The algorithm is straightforward.  For each detector it combines the
points from each method according to the matching and global count
criteria, to get a per-detector changepoint list in step~2.  The
method's returned value may need post-processing to convert its
result into points. Step~2b1 drops the method if it is noisy.
Steps~3 and 4 determine the consistency bounds on the per-library
point count, and step~5 drops those libraries whose point counts
fall outside.  Step~6 forms a master changepoint list, which is
screened in step~8 to keep only those points matching a detector's
result in a majority of cases.

\func{remove\_nearby} replaces points within $ \Delta $ indices of each
other with a single point at their center of mass.  It uses single
linkage that can chain across the data.  \func{mean} calculates the
average of the values passed.

\begin{center}
\begin{algo}{algo:cpt}
\alghdr{{\bf cpt}: Fusion of Changepoint Detectors} \\
\algrow[Input:]{ signal $ x $, changepoint detectors $ \sets{C} $
                   with methods $ M_{\sets{C}} $, }
\algrow{ maximum changepoint fraction $ f_{pt} $,
         point count range $ q_{vote} $, }
\algrow{ maximum separation relative $ f_{\Delta} $, absolute $ \Delta_{max} $, }
\algrow{ intra-library separation $ \Delta_{lib} $ }
\algrow[1:]{ let merge distance $ \Delta \leftarrow
               \min(\text{round}(f_{\Delta} \sizeof{x}),\, \Delta_{max}) $ }
\algrow[2:]{ for each detector $ C_{i} $ in $ \sets{C} $ }
\algrowi[2a:]{ create array of methods point lists
                 $ p_{raw}[1 \ldots \sizeof{C_{i}},
                           1 \ldots \sizeof{M_{C_{i}}}] $ }
\algrowi[2b:]{ for each method $ M_{j} $ in $ C_{i} $ }
\algrowii[2b1:]{ $ p_{raw}[i,j] \leftarrow M_{j}(x) $ }
\algrowii{ if $ f_{pt} \sizeof{x} \leq \sizeof{p_{raw}[i,j]} $ then }
\algrowiii[2b2a:]{ $ p_{raw}[i,j] \leftarrow \varnothing $ }
\algrowi[2c:]{ let library points list
  $ p[i] \leftarrow
    \text{remove\_nearby}(p_{raw}[i,1 \ldots j], \Delta_{lib}) $ }
\algrow[3:]{ let library points count
  $ np[1 \ldots \sizeof{\sets{C}}] \leftarrow
    \sizeof{p[1 \ldots \sizeof{\sets{C}}]} $ }
\algrow[4:]{ let allowed point count $ N_{pt} \leftarrow
                \text{quantile}(np, q_{vote}) $ }
\algrow[5:]{ for $ i $ in $ 1 \ldots \sizeof{\sets{C}} $ }
\algrowi{ if $ np[i] \leq N_{pt}[1] \text{ OR } np[i] \geq N_{pt}[2] $ }
\algrowii[5a1:]{ $ p[i] \leftarrow \varnothing $ }
\algrow[6:]{ $ p_{common} \leftarrow \text{remove\_nearby}(p, \Delta) $ }
\algrow[7:]{ let remaining library count $ N_{lib} \leftarrow
                \#\{ p \neq \varnothing \} $ }
\algrow[8:]{ for $ i $ in $ 1 \ldots | p_{common} | $ }
\algrowi{ if $ N_{lib} / 2 \leq \#\{min(p_{common}[i] - p) \leq \Delta\} $
               then }
\algrowii[8a1:]{ report $ p_{common}[i] $ }
\end{algo}
\end{center}

\begin{center}
\begin{algo}{algo:rmnear}
\alghdr{{\bf remove\_nearby}: Point Simplification} \\
\algrow[Input:]{ points list $ p $, maximum separation $ \Delta $ }
\algrow{ if $ \sizeof{p} = 1 $ then }
\algrowi[1a:]{ return $ p $ }
\algrow[2:]{ $ p \leftarrow \text{sort}(p, \text{INCREASING}) $ }
\algrow[3:]{ for $ i $ in $ 2 \ldots |p| $ }
\algrowi[3a:]{ $ j \leftarrow i $ }
\algrowi[3b:]{ while $ j \leq \sizeof{p}
                  \text{ AND } (p[j] - p[j-1]) < \Delta $ }
\algrowii[3b1:]{ $ j \leftarrow j + 1 $ }
\algrowi[3c:]{ $ j \leftarrow j - 1 $ }
\algrowi{ if $ i \leq j $ then }
\algrowii[3d1:]{ $ p[i] \leftarrow \text{mean}(p[i \ldots j]) $ }
\algrowii[3d2:]{ delete $ p[(i+1) \ldots j] $ }
\algrow[4:]{ return $ p $ }
\end{algo}
\end{center}

\end{appendix}

\end{document}